\documentclass{PoS}

\title{The PHENIX Potential in the Search for the QCD Critical Point}

\ShortTitle{PHENIX and the Critical Point}

\author{\speaker{J.T. Mitchell}\thanks{for the PHENIX Collaboration.}\\
        Brookhaven National Laboratory\\
        E-mail: \email{mitchell@bnl.gov}}

\abstract{
With the measurement of several observables at SPS energies that demonstrate non-monotonic behavior as a function of centrality and $\sqrt{s_{NN}}$, there is growing interest in pursuing a scan of relativistic heavy ion collisions at low energies at the Relativistic Heavy Ion Collider. The capabilities of the PHENIX experiment to take quality measurements at low RHIC energies is described and directly demonstrated with analyses of Au+Au collisions at $\sqrt{s_{NN}}$ = 19.6 GeV and Cu+Cu collisions at $\sqrt{s_{NN}}$ = 22.5 GeV. The contribution of upgrades to the PHENIX detector in the upcoming years will also be discussed in the context of a low energy RHIC run.
         }

\FullConference{The 3rd edition of the International Workshop --- The Critical Point and Onset of Deconfinement --- \\
		 July 3-7 2006\\
		 Galileo Galilei Institute, Florence, Italy}

\begin{document}

\section{Introduction}

Data from relativistic heavy ion collisions with collision energies between that provided by the AGS and SPS accelerators ($\sqrt{s_{NN}}$ = 5-10 GeV) provide hints that the onset of critical behavior may be occuring in that region \cite{SPS}. To date, many of these measurements have yet to be corraborated by more than one experiment. Since the versatile RHIC accelerator can provide collisions in the region of interest, the RHIC experiments have a unique opportunity to help provide a wealth of information concerning the location of the QCD critical point, and this data can be taken with high statistics in multiple complementary experiments.

The PHENIX Experiment at RHIC is designed as a versatile detector that can make quality measurements of both global properties and rare probes from elementary and heavy ion collisions. Details of the PHENIX detector set-up can be found elsewhere \cite{phenixNIM}. In this article, it will be demonstrated that PHENIX is currently capable of making measurements at low beam energies by reviewing several recent results from 19.6 GeV Au+Au and 22.5 GeV Cu+Cu collisions. The PHENIX capabilities will become more comprehensive as the upcoming set of detector upgrades to PHENIX are implemented.

\section{Low Energy Performance and Recent Results}

The PHENIX detector, like the RHIC accelerator, has a very versatile design that allows it to make quality measurements over a very wide range of collision energies and species with little or no degradation in performance.  The PHENIX data acquisition system (DAQ) \cite{phenixDAQ} is capable of taking data at a very high rate that will easily handle the estimated RHIC luminosity at low energy.

\begin{figure}
\resizebox{0.5\textwidth}{!}{%
  \includegraphics{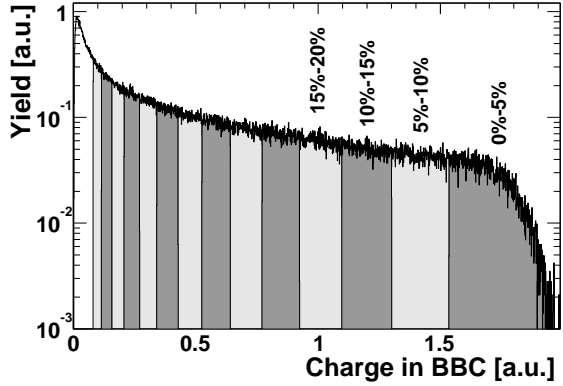}
}
\caption{The yield of the total charge deposited in the Beam-Beam Counters for minimum bias 200 GeV Au+Au collisions. Both the charge and the yield are displayed in arbitrary units.  The shaded regions indicate increments of 5\% of the total integrated yield.}
\label{fig:bbcCentrality}
\end{figure}

One capability that suffers a performance degradation at low collision energies is the determination of the centrality of the collision.  This is due to the fixed location of the Beam-Beam Counters (BBC) \cite{bbcNIM}, which are the primary detectors used for centrality determination in collisions at 200 and 62 GeV, as shown in Fig. \ref{fig:bbcCentrality}. As the beam energy decreases, especially below $\sqrt{s_{NN}} \approx 50$ GeV, it becomes kinematically possible for spectator nucleons to fall within the acceptance of the BBC ($3.0<|\eta|<4.0$), which affects the linearity of the BBC response as a function of the number of participating nucleons, $N_{part}$ or $N_{p}$. This problem can be overcome by using the existing PHENIX central arm spectrometer, which can measure the charged particle multiplicity near mid-rapidity in Pad Chamber 1 (PC1) \cite{centralNIM}. This solution has the drawback that it can introduce autocorrelations into some measurements, such as the measurement of elliptic flow, that directly use the PHENIX central arm.  A detailed description of the procedure used to define centrality in 19.6 GeV Au+Au collisions using the PHENIX central arm can be found elsewhere \cite{ppg019}. The PHENIX detector could benefit from the addition of a new detector dedicated to measuring centrality at low beam energies.

The most convincing way to demonstrate the current PHENIX capability for making measurements at low beam energy is to review results taken during the short RHIC runs near SPS energies in RHIC Run-2 (19.6 GeV Au+Au, 40,000 minimum bias events, from one day of data taking with no magnetic field in PHENIX) and Run-5 (22.5 GeV Cu+Cu, 9 million minimum bias events, also collected in about one day).  These results each use the total PC1 charged particle multiplicity as a centrality measure.

\begin{figure}
\resizebox{0.5\textwidth}{!}{%
  \includegraphics{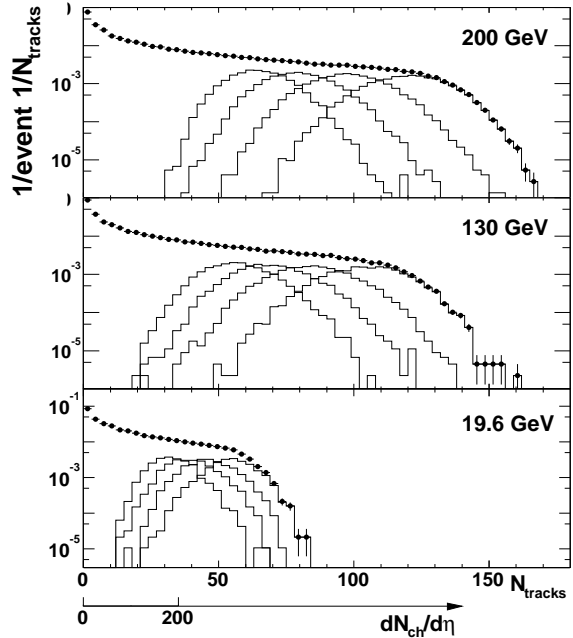}
}
\caption{The distribution of the number of tracks in the east arm of the PHENIX detector per minimum bias trigger, measured at three energies. The axis on the bottom corresponds to mid-rapidity values of $dN_{ch}/d\eta$. Distributions of the four 5\% most central bins are also shown in each plot.}
\label{fig:nch}
\end{figure}

\begin{figure}
\resizebox{0.5\textwidth}{!}{%
  \includegraphics{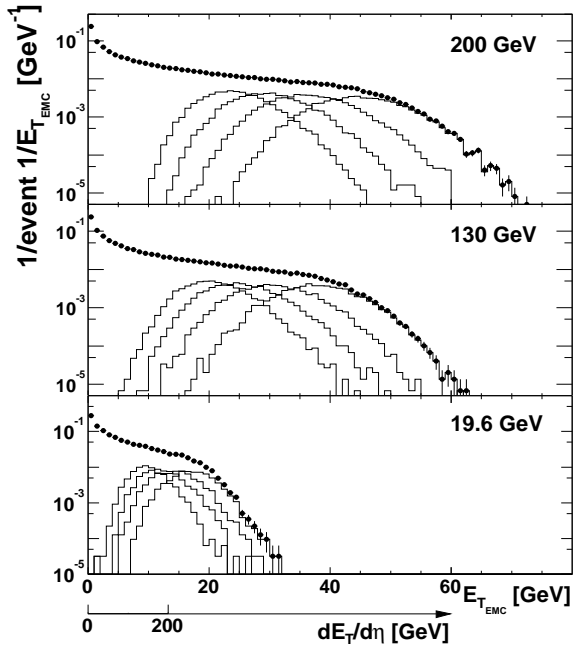}
}
\caption{The distribution of the raw transverse energy in two calorimeter sectors of the PHENIX detector per minimum bias trigger, measured at three energies. The axis on the bottom corresponds to mid-rapidity values of $dE_{T}/d\eta$. Distributions of the four 5\% most central bins are also shown in each plot.}
\label{fig:et}
\end{figure}

\begin{figure}
\resizebox{0.5\textwidth}{!}{%
  \includegraphics{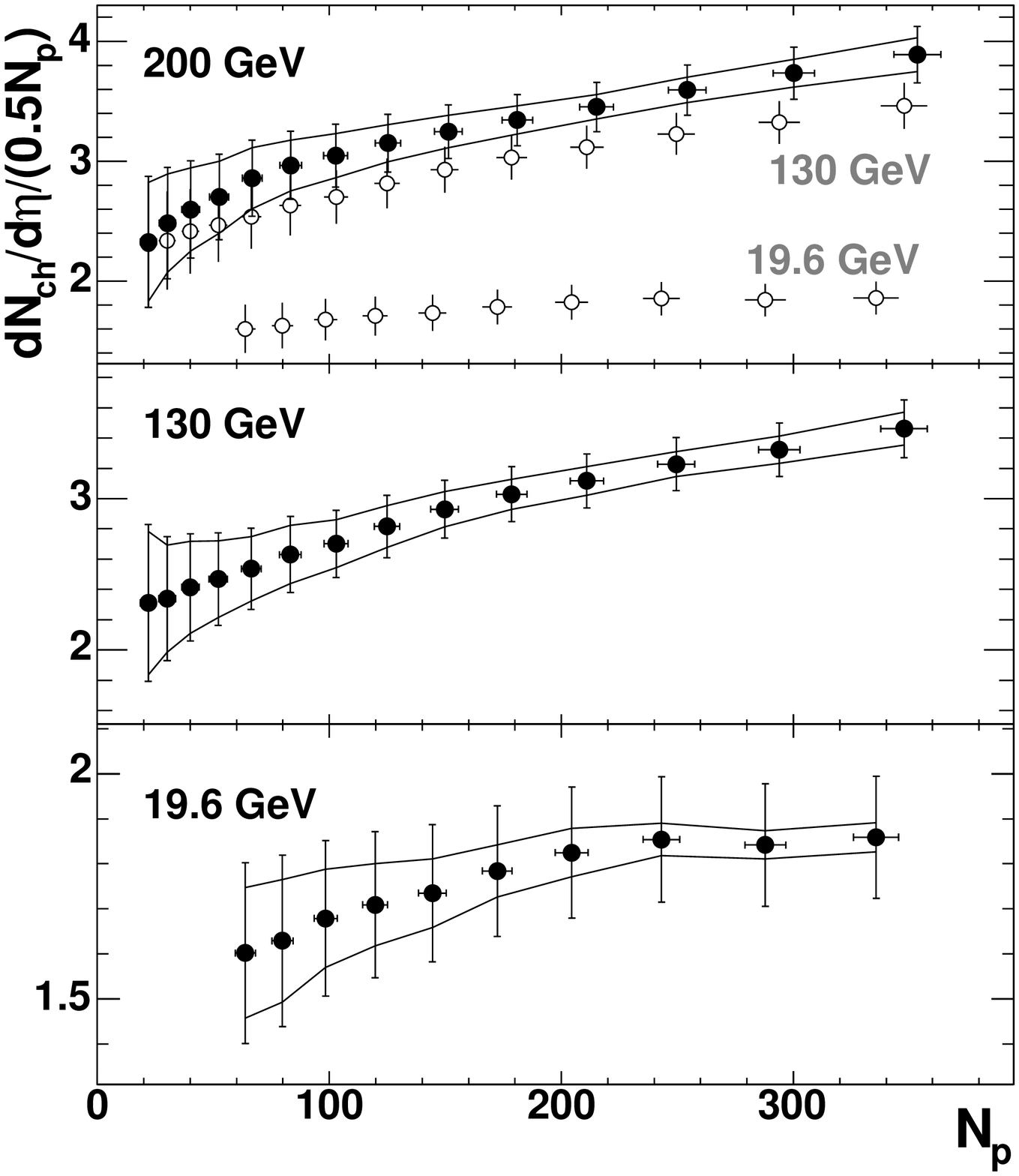}
}
\caption{$dN_{ch}/d\eta$ divided by the number of participant pairs at three RHIC energies. Errors shown with vertical bars are systematic errors. Lines show the part of the systematic error that allows for bending or inclination of the point. Horizontal errors denote the uncertainty in the determination of $N_{part}$.}
\label{fig:nchNp}
\end{figure}

\begin{figure}
\resizebox{0.5\textwidth}{!}{%
  \includegraphics{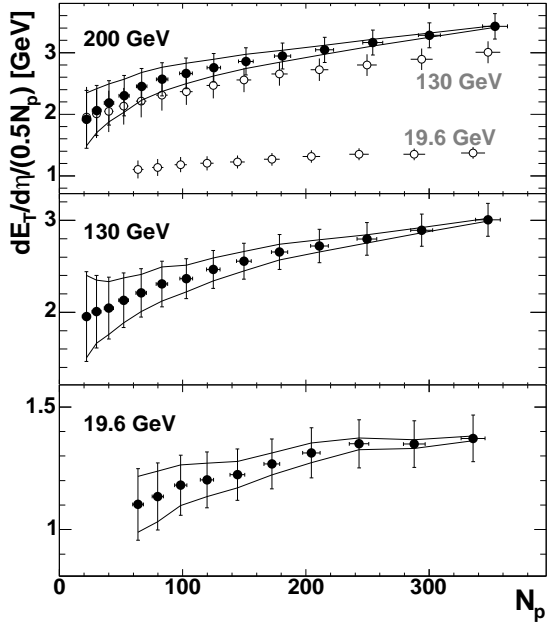}
}
\caption{$dE_{T}/d\eta$ divided by the number of participant pairs at three RHIC energies. Errors shown with vertical bars are systematic errors. Lines show the part of the systematic error that allows for bending or inclination of the point. Horizontal errors denote the uncertainty in the determination of $N_{part}$.}
\label{fig:etNp}
\end{figure}

PHENIX has published measurements of inclusive charged particle multiplicity and total transverse energy in 19.6 GeV Au+Au collisions \cite{ppg019}. Fig. \ref{fig:nch} and \ref{fig:et} show the $dN_{charged}/d\eta$ and $dE_{T}/d\eta$ distributions as a function of centrality, which exhibit the characteristic knee shape for these measurements. These measurements have been normalized by the number of participant pairs in  Fig. \ref{fig:nchNp} and \ref{fig:etNp}. The results are consistent with SPS results \cite{ppg019} and are consistent with the linear scaling observed for $dE_{T}/d\eta/(0.5 N_{part}$) and $dN_{charged}/d\eta/(0.5 N_{part}$) as a function of $\sqrt{s_{NN}}$ in Au+Au collisions \cite{ppg019}.

The PHENIX detector can identify particles using time-of-flight with high resolution in both the time-of-flight (TOF) detectors \cite{phenixTOF} and the electromagnetic calorimeters (EMC) \cite{phenixEMC}. The separation of kaons and pions has been achieved in the TOF for $p_T > 3$ GeV/c and in the EMC for $p_T > 1$ GeV/c. At low energy, the primary factor degrading particle identification performance is the resolution of locating the collision vertex along the beam axis due to lower particle multiplicities in the BBC, which in turn affects the momentum resolution. Fortunately, track projections in the central arm can also be used to measure the collision vertex on an event-by-event basis with minimal performance degradation.  In 22.5 GeV Cu+Cu collisions, the particle identification performance is at or near that at higher energies.

\begin{figure}
\resizebox{0.5\textwidth}{!}{%
  \includegraphics{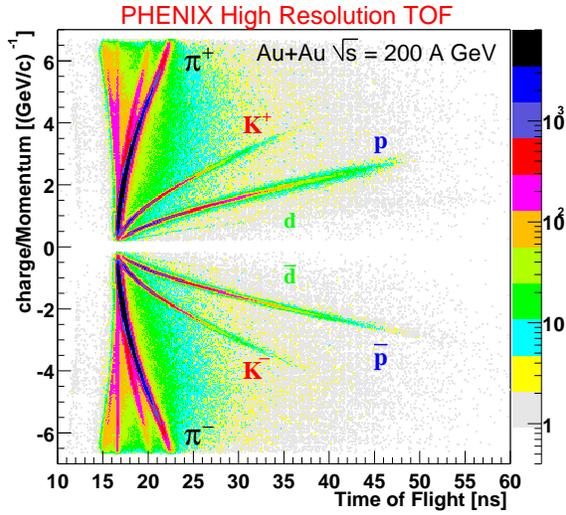}
}
\caption{The charge divided by the momentum versus the time-of-flight measured by the TOF detector. Clear bands corresponding to pions, kaons, and protons are apparent.}
\label{fig:tofPid}
\end{figure}

\begin{figure}
\resizebox{0.5\textwidth}{!}{%
  \includegraphics{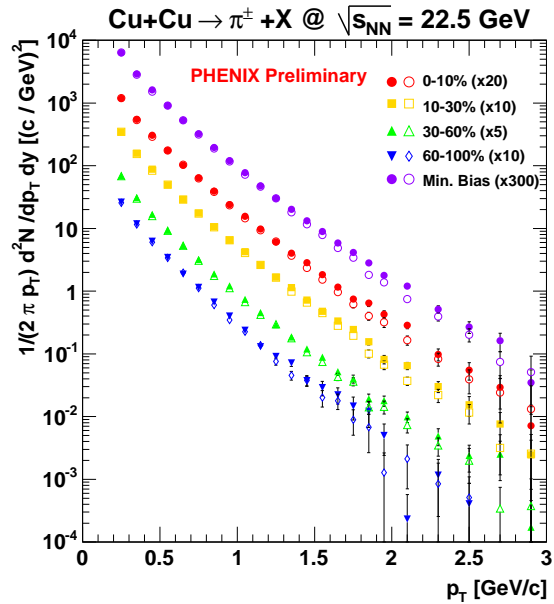}
}
\caption{PHENIX Preliminary transverse momentum spectra for pions from 22.5 GeV Cu+Cu collisions. The spectra are shown for various centrality bins, with each scaled by the amount indicated in the legend.}
\label{fig:pionSpectra}
\end{figure}

\begin{figure}
\resizebox{0.5\textwidth}{!}{%
  \includegraphics{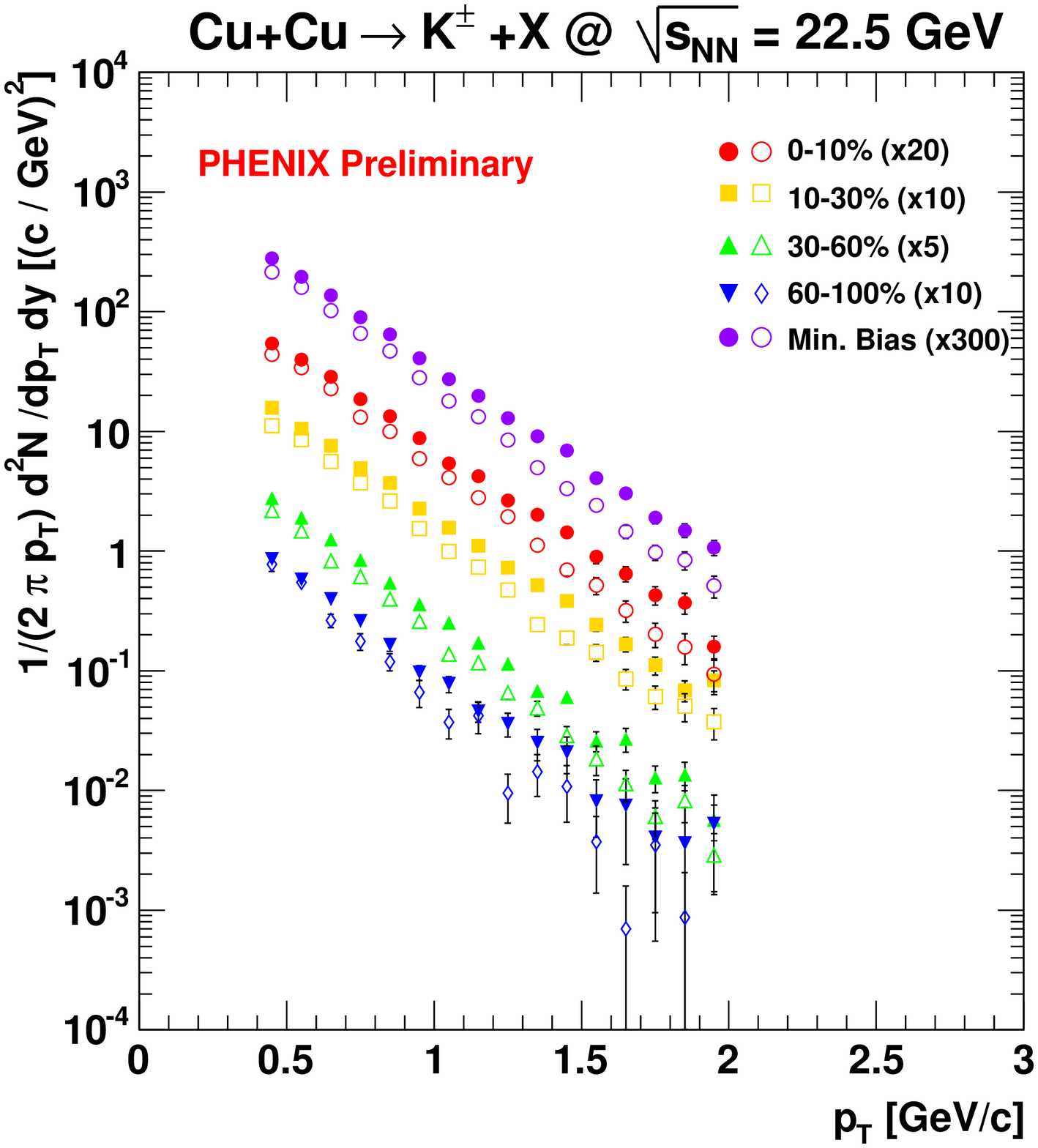}
}
\caption{PHENIX Preliminary transverse momentum spectra for kaons from 22.5 GeV Cu+Cu collisions. The spectra are shown for various centrality bins, with each scaled by the amount indicated in the legend.}
\label{fig:kaonSpectra}
\end{figure}

\begin{figure}
\resizebox{0.5\textwidth}{!}{%
  \includegraphics{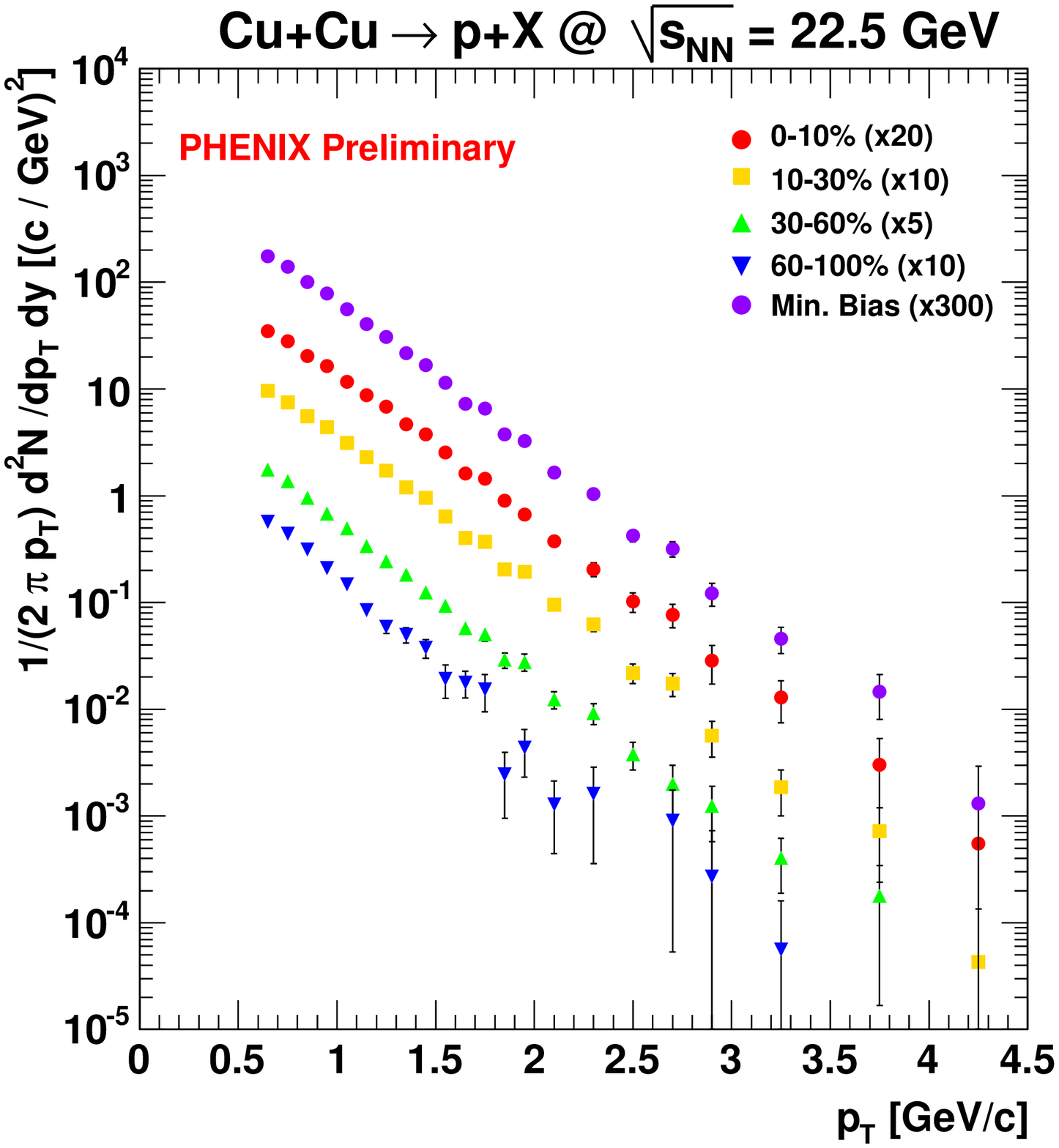}
}
\caption{PHENIX Preliminary transverse momentum spectra for protons from 22.5 GeV Cu+Cu collisions. The spectra are shown for various centrality bins, with each scaled by the amount indicated in the legend.}
\label{fig:protonSpectra}
\end{figure}

\begin{figure}
\resizebox{0.5\textwidth}{!}{%
  \includegraphics{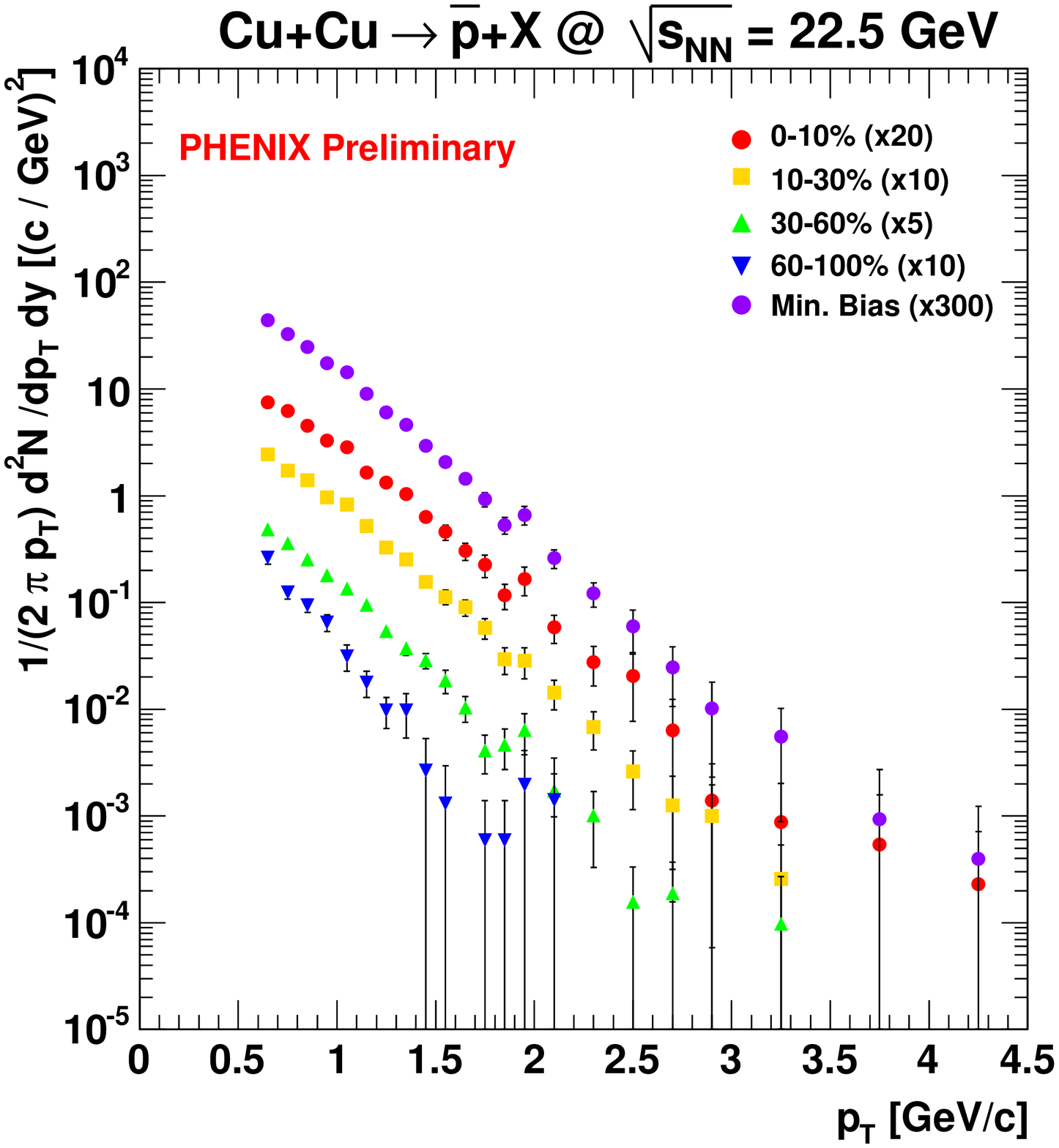}
}
\caption{PHENIX Preliminary transverse momentum spectra for anti-protons from 22.5 GeV Cu+Cu collisions. The spectra are shown for various centrality bins, with each scaled by the amount indicated in the legend.}
\label{fig:pbarSpectra}
\end{figure}

\begin{figure}
\resizebox{0.5\textwidth}{!}{%
  \includegraphics{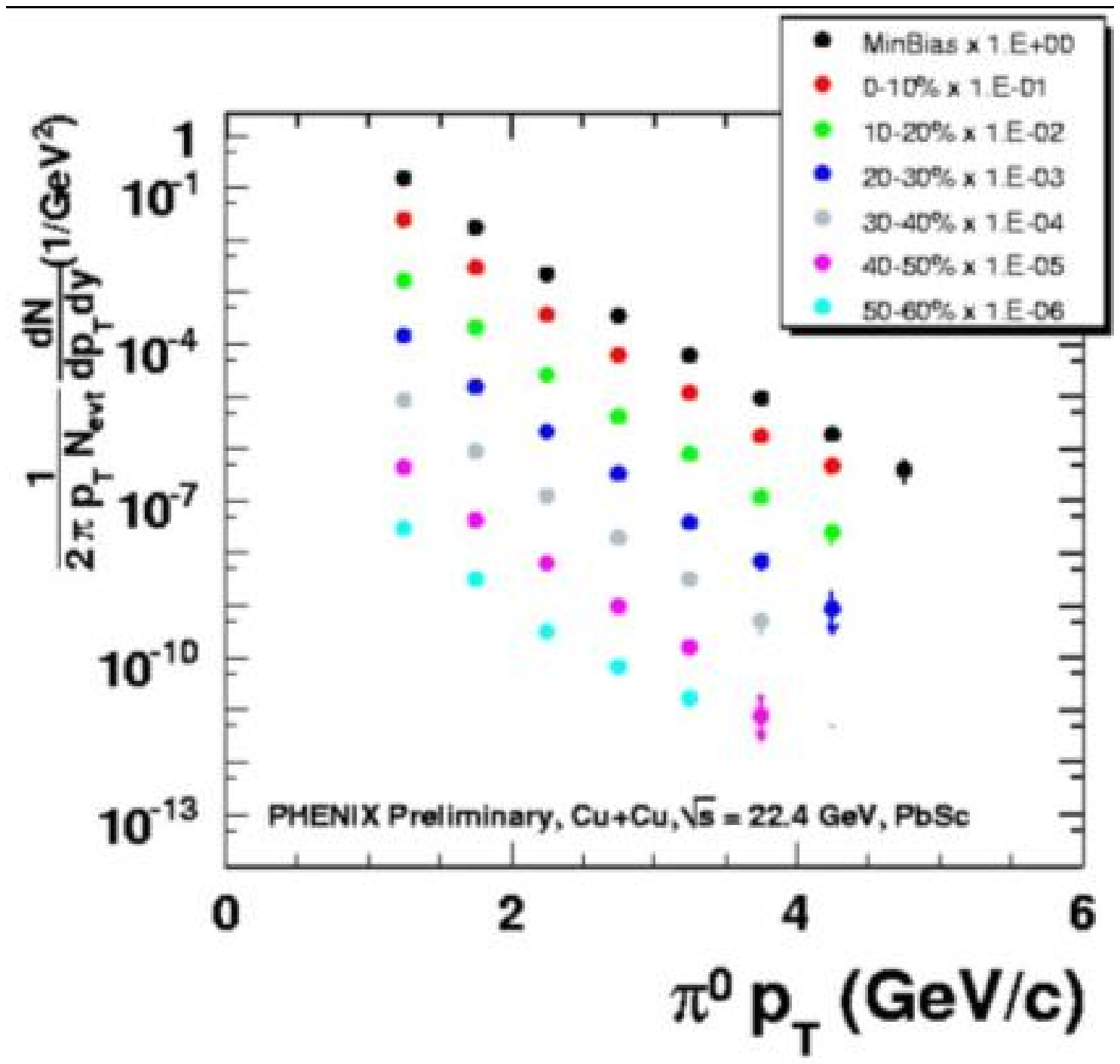}
}
\caption{PHENIX Preliminary transverse momentum spectra for neutral pions from 22.5 GeV Cu+Cu collisions. The spectra are shown for various centrality bins, with each scaled by the amount indicated in the legend.}
\label{fig:pi0Spectra}
\end{figure}

PHENIX has utilized the TOF particle identification capability to analyze transverse momentum spectra of pions, protons, and kaons in 22.5 GeV Cu+Cu collisions. Figure \ref{fig:pionSpectra} shows the $p_T$ spectra for positive and negative pions for various centrality selections.  For this dataset, which was taken in roughly a single day of RHIC running, the pion spectra have been measured to a $p_T$ of 3 GeV/c. In addition, figure \ref{fig:kaonSpectra} shows kaon $p_T$ measured to 2 GeV/c, and figures \ref{fig:protonSpectra} and \ref{fig:pbarSpectra} show proton and antiproton $p_T$ measured to 3 GeV/c. Also shown in figure \ref{fig:pi0Spectra} are neutral pion spectra using the EMC measured to $p_T$ of about 4 GeV/c These spectra demonstrate that enough statistics for a precise determination of the freeze-out temperature can be accumulated in a relatively short amount of time.

\begin{figure}
\resizebox{0.5\textwidth}{!}{%
  \includegraphics{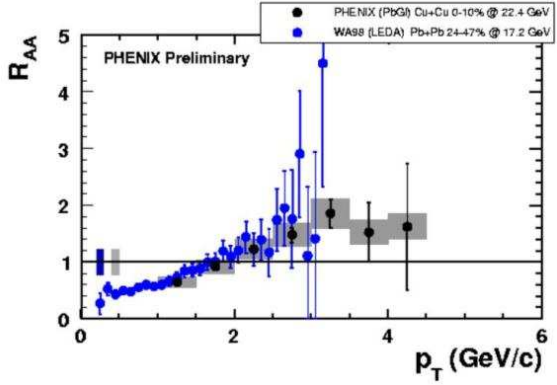}
}
\caption{PHENIX Preliminary $R_{AA}$ as a function of $p_{T}$ for neutral pions from 0-10\% central 22.5 GeV Cu+Cu collisions. Overlayed is $R_{AA}$ for neutral pions measured by WA98 at the SPS in 24-47\% central 17 GeV Pb+Pb collisions.}
\label{fig:p0Raa}
\end{figure}

\begin{figure}
\resizebox{0.5\textwidth}{!}{%
  \includegraphics{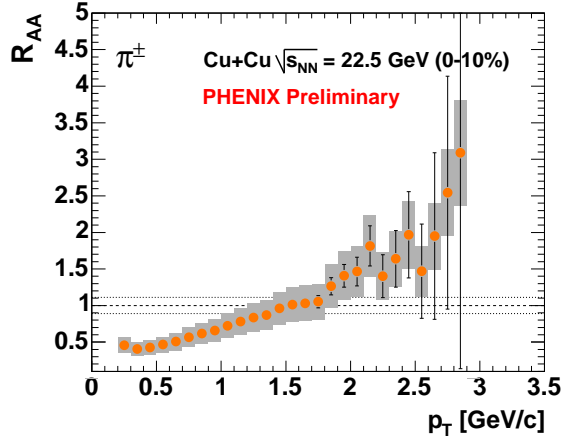}
}
\caption{PHENIX Preliminary $R_{AA}$ as a function of $p_{T}$ for charged pions from 0-10\% central 22.5 GeV Cu+Cu collisions.}
\label{fig:pionRaa}
\end{figure}

An observable of particular interest is that of the nuclear modification factor at high $p_T$, $R_{AA}$.  $R_{AA}$ is defined as
\begin{equation}
R_{AA} = \frac{d^{2}N_{AA}/dp_{T}dy}{<T_{AA}>d^{2}\sigma_{pp}/dp_{T}dy},
\end{equation}
where $<T_{AA}>$ is the Glauber nuclear overlap function. The value of $R_{AA}$ has been measured by PHENIX for 22.5 GeV Cu+Cu collisions, and is shown for neutral pions in figure \ref{fig:p0Raa} and charged pions in figure \ref{fig:pionRaa}.  From these figures, it is apparent that at 22.5 GeV, the suppression observed at higher collision energies is no longer present.  The precision of the measurement of $R_{AA}$, among other observables, depends on the precision of the baseline p+p spectra.  In order to reduce systematic uncertainties on low energy measurements, it is highly desireable to obtain companion p+p and/or d+Au spectra at each energy setting.

\begin{figure}
\resizebox{0.5\textwidth}{!}{%
  \includegraphics{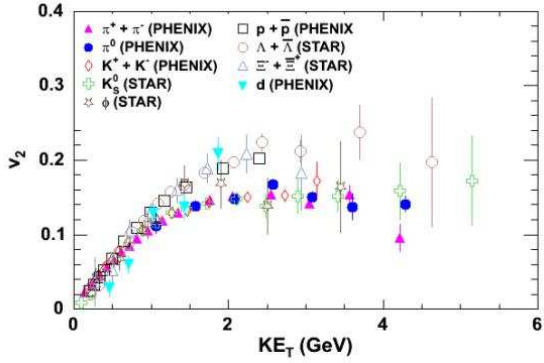}
}
\caption{Elliptic flow in minimum bias 200 GeV Au+Au collisions in terms of $v_2$ for various identified particles as a function of transverse kinetic energy, showing a separate universal scaling behavior for mesons and baryons.}
\label{fig:v2}
\end{figure}

PHENIX has demonstrated that elliptic flow measurements ($v_2$) of identified particles in Au+Au collisions exhibit universal scaling as a function of transverse kinetic energy, $KE_T = m_T - m$, as shown in Fig. \ref{fig:v2} \cite{phenixV2}. This scaling is distinct for mesons and baryons, indicating that quark degrees of freedom may be driving the flow. In low energy RHIC running, PHENIX will be able to test where these scaling properties break down, which may also provide further information on the location of the critical point.

\section{Searching for Signals of Critical Behavior}

Fluctuation measurements remain a powerful tool to be utilized in the search for direct evidence of the phase transition and the location of the critical point. A low energy scan at RHIC would open up the possibility to measure several critical exponents \cite{Stanley} in heavy ion collisions.  Near the critical point, several observables of a system are expected to diverge.  Some of these observables include the heat capacity, the compressibility, and the correlation length.  These properties can be probed with transverse momentum fluctuations, multiplicity fluctuations, and two-particle correlations, respectively.  The rate of divergence of these quantities near the critical point for a second-order phase transition is described by power laws with a set of critical exponents. For all systems within the same universality class, the exponents are identical. Hence, critical behavior is manifested by universal power law behavior as a function of the temperature of the system.

An example of how a critical exponent analysis may be performed can be made with current charged particle multiplicity fluctuations data from PHENIX.  In the Grand Canonical Ensemble, fluctuations in the particle number can be directly related to the compressibility as follows:
\begin{equation}
\sigma^2/\mu^2 = k_B \frac{T}{V} k_T,
\end{equation}
where $k_B$ is the Boltzmann constant, $k_T$ is the isothermal compressibility, T is the temperature, and V is the volume \cite{Stanley}. At the critical point, the divergence of the compressibility is described by the critical exponent $\gamma$ as follows:
\begin{equation}
k_T = A ((T-T_C)/T_C)^{-\gamma},
\end{equation}
where $T_C$ is the critical temperature \cite{Stanley} and $A$ is a constant.

\begin{figure}
\resizebox{0.5\textwidth}{!}{%
  \includegraphics{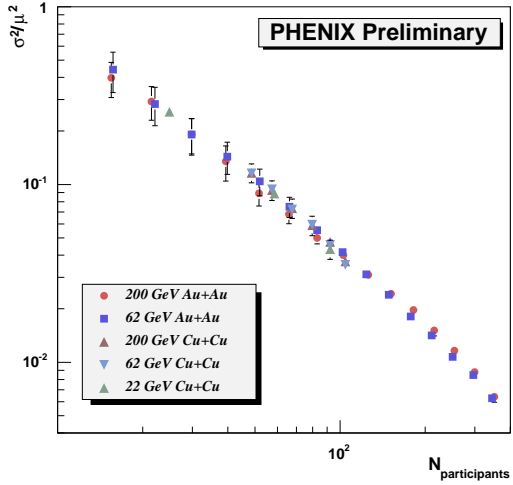}
}
\caption{Dynamical multiplicity fluctuations in terms of $\sigma^2/\mu^2$ for several systems including 22.5 GeV Cu+Cu, scaled to the 200 GeV Au+Au data to emphasize the power law scaling with similar exponents for all systems.}
\label{fig:svarMu2}
\end{figure}

Fluctuations in terms of $\sigma^2/\mu^2$ are shown in Figure \ref{fig:svarMu2} for five different collision systems at 3 beam energies. These fluctuations have been corrected to remove non-dynamical impact parameter fluctuations and have been extrapolated to 2$\pi$ acceptance. All species have been scaled to the 200 GeV Au+Au data in order to best illustrate the fact that all systems exhibit a universal power law scaling behavior as a function of $N_{part}$.  All datasets can be described by the curve $\sigma^2/\mu^2 \propto N_{part}^{-1.40 \pm 0.03}$. However, since it is unlikely that there is a significant temperature variation as a function of $N_{part}$ as evidenced by estimates of the centrality-dependence of the freeze-out temperature from PHENIX data, it is not prudent to interpret this data as evidence for critical behavior in these systems. However, in a RHIC energy scan, it will be possible to measue multiplicity fluctuations as a function of $\sqrt{s_{NN}}$ in the region of interest and then extract the the value of $\gamma$ using the freeze-out temperature estimated from the identified particle spectra and Hanbury-Brown Twiss correlation data as described in \cite{phenixPhi}. If QCD belongs in a universality class with identical exponents as the mean field theory \cite{qcdMeanField}, then a value for the critical exponent $\gamma$ of 1.0 should be extracted. The simultaneous measurement of several critical exponents in this way could provide unambiguous evidence of the existence of the QCD critical point.

\section{Future Prospects}

PHENIX has several upgrades planned for installation in the near future that will greatly strengthen the PHENIX capabilities for low energy running.  These include a forward muon trigger detector, a silicon vertex tracking detector for precision vertex tracking, an Aerogel Cerenkov Counter, a multi-gap resistive plate chamber time-of-flight detector that will extend the PHENIX particle identification capabilities for pion, kaon, and proton separation to transverse momenta up to 10 GeV/c, a hadron blind detector that will greatly improve rejection of Dalitz and conversion electrons, a nose cone calorimeter and muon piston calorimeter that will provide photon and $\pi^{0}$ coverage to forward rapidities, a dedicated reaction plane detector that will improve the reaction plane resolution by a factor of two, and an upgrade to the Data Acquisition system (DAQ).  All of these upgrades will be available for the RHIC run scheduled for 2006-2007 with the exception of the silicon vertex tracker, the nose cone calorimeter, and the DAQ upgrade, which are scheduled for installation in 2009, 2011, and 2011, respectively.

The reaction plane detector, which consists of scintillator paddles with lead converters located at $1<|\eta|<3$, will be available for the next RHIC running period. This detector will prove very useful for low energy running since it is capable of serving as a trigger counter due to its location away from the beam fragmentation region. It may also have the capability to better define collision centrality and event vertex when compared to the resolution offered by PC1. The installation of the silicon vertex tracker, which covers $2\pi$ in azimuth for $|\eta|<1.2$ will vastly improve the centrality and event vertex resolution \cite{phenixSVX}. Although this detector is designed for detecting displaced vertices from the decay of mesons containing charm or bottom quarks, it is also an ideal detector for measuring charged particle multiplicity, multiplicity fluctuations, and flow over a much improved acceptance when compared to the current capabilities of the PHENIX central arm.

\section{Conclusion}

This article has outlined the present and future capabilities of the PHENIX detector for low energy RHIC running. With the current detector and the set of upgrades scheduled to be installed over upcoming RHIC runs, PHENIX is a constantly improving tool that is ready to take quality data in low energy RHIC running configurations. PHENIX is prepared to measure a wide variety of complementary observables that should be able to provide unambiguous signals of the location of the QCD critical point. Measurements with the necessary precision can be accumulated over a few days of RHIC running at each energy setting. It is also imperative that partner baseline p+p and/or d+Au datasets be accumulated at each setting to serve as a control and greatly improve the precision of measurements such as $R_{AA}$. Comprehensive critical exponent analyses would benefit from energy scans over more than one heavy ion species.

\end{document}